\documentclass[prl,preprint,endfloats,amsmath,amssymb]{revtex4}
\usepackage{graphicx}
\usepackage{dcolumn}
\usepackage{bm}

\begin{document}

\newcommand{\be}{\begin{equation}}
\newcommand{\ee}{\end{equation}}
\newcommand{\bes}{\begin{equation}}
\newcommand{\ees}{\nonumber\end{equation}}
\newcommand{\bea}{\begin{eqnarray}}
\newcommand{\eea}{\end{eqnarray}}
\newcommand{\uv}[1]{\mathbf{\hat{#1}}}
\newcommand{\curl}[1]
{\mathbf{\nabla}\times\mathbf{#1}}
\newcommand{\dvg}[1]
{\mathbf{\nabla}\cdot\mathbf{#1}}
\newcommand{\pd}[2]
{{{\partial #1} \over {\partial #2}}}
\newcommand{\pdt}[2]
{{{\partial^{2} #1} \over {\partial #2 ^{2}}}}

\newcommand{\m}[1]
{\mathbf{#1}}


\title{Thermal-magnetic noise measurement of spin-torque effects on ferromagnetic resonance in MgO-based magnetic tunnel junctions}

\author{Y. Guan\footnote{The author is currently with SoloPower Inc., 5981 Optical Court, San Jose, California 95138. Electronic mail: yguan@solopower.com.} and J. Z. Sun}
 \affiliation{IBM T. J. Watson Research Center, Yorktown Heights, New York 10598, USA}
\author{X. Jiang, R. Moriya, L. Gao, and S. S. P. Parkin}
 \affiliation{IBM Almaden Research Center, San Jose, California 95120, USA}

\date{\today}

\begin{abstract}

Thermal-magnetic noise at ferromagnetic resonance (T-FMR) can be
used to measure magnetic perpendicular anisotropy of nanoscale
magnetic tunnel junctions (MTJs). For this purpose, T-FMR
measurements were conducted with an external magnetic field up to 14
kOe applied perpendicular to the film surface of MgO-based MTJs
under a dc bias. The observed frequency-field relationship suggests
that a 20 $\mathring{\textrm{A}}$ CoFeB free layer has an effective
demagnetization field much smaller than the intrinsic bulk value of
CoFeB, with 4$\pi M_{eff}$ = (6.1 $\pm$ 0.3) kOe. This value is
consistent with the saturation field obtained from magnetometry
measurements on extended films of the same CoFeB thickness. In-plane
T-FMR on the other hand shows less consistent results for the
effective demagnetization field, presumably due to excitations of
more complex modes. These experiments suggest that the perpendicular
T-FMR is preferred for quantitative magnetic characterization of
nanoscale MTJs.

\end{abstract}


\maketitle

Understanding and controlling magnetic properties of thin layers in
a spatially confined magnetic system are of growing interest for
fundamental physics studies and for spin-torque-based magnetic
memory and microwave oscillator applications\cite{Sankey-2006,
Fuchs-2007, Chen-2008, Houssameddine-2008, Deac-2008}. Specifically,
the layer magnetic anisotropy, damping, and effective magnetization,
are of particular importance for the performance of
spin-torque-based devices\cite{Sun-2000}. Although conventional
ferromagnetic resonance (FMR) detection methods lack the sensitivity
to measure individual nanoscale structures, thermal-magnetic noise
measurement of ferromagnetic resonance (T-FMR) enables direct
studies of magnetic properties of patterned nanoscale
devices\cite{Nazarov-2002, Synogatch-2003, Stutzke-2003, Petit-2007,
Petit-2008, Guan-2009}.

For a dc-biased nanoscale magnetic device below its
spin-torque-induced magnetic instability, the system is essentially
driven by thermal noise. By monitoring high-frequency thermal
magnetization fluctuations of the device, field- and bias-dependent
T-FMR spectra can be obtained. Thermal fluctuations play a
significant role in determining spin-torque-driven magnetic
switching in nanoscale magnetic tunnel junctions
(MTJs)\cite{Devolder-2008}. Therefore, T-FMR could also be useful
for further understanding spin-torque dynamics in MTJs.

In this letter, we present room-temperature T-FMR studies of
MgO-based nanopillar MTJs between 2 and 8 GHz, where an external
magnetic field up to 14 kOe is applied perpendicular to the film
surface of the subcritically dc-biased MTJs. The observed
perpendicular field dependence of the T-FMR frequency is consistent
with Kittel formula. The effective demagnetization field of the
CoFeB free layer has been determined, which is compared with those
obtained on extended films of the same CoFeB thickness.

The nanopillar MTJs have a stack structure of
Si/SiO$_{2}$/Ta(75)/Cu(200)/Ta(50)/
IrMn(120)/CoFeB(6)/CoFe(30)/Ru(7)/CoFe(27)/MgO(10)/CoFeB(20)/Ta(50)/Ru(50),
where the numbers are layer thickness in angstroms, CoFe =
Co$_{70}$Fe$_{30}$, and CoFeB = Co$_{40}$Fe$_{40}$B$_{20}$.
Nanopillars were patterned down to about 100 nm using electron beam
lithography. The main results presented here are from a
representative 50 $\times$ 140 nm$^{2}$ MTJ with a barrier
resistance-area product (RA) of $\sim$11 $\Omega$ $\mu m^{2}$ and a
tunneling magnetoresistance ratio (TMR) of $\sim$110$\%$. The MTJ
was fully patterned through the pinned layer, stopping at the bottom
IrMn.

A simplified diagram of the T-FMR experimental setup is shown in
Fig. 1(a), where a 50 $\Omega$ bias-T is used for simultaneous dc
and microwave signal coupling into and out of the MTJ. The dc bias
current ($I$) is defined positive when current flows from the top
free layer (FL) to the bottom pinned layer (PL). The output T-FMR
signals are fed into a 35 dB low-noise amplifier and then into a
spectrum analyzer. The resolution bandwidth of the spectrum analyzer
is set to 5 MHz. Each spectrum is obtained by averaging over 100
spectral scans. A more detailed description about the T-FMR
experimental setup can be found in our previous
report\cite{Guan-2009}.

Perpendicular T-FMR measurements are performed at room temperature
with an external magnetic field up to 14 kOe applied perpendicular
to the film surface of the subcritically dc-biased MTJ up to 360
$\mu$A. Figure 1(b) shows the tunnel resistance ($R$) of the MTJ as
a function of the applied perpendicular magnetic field ($H$) for a
small dc bias of $I$ = 1 $\mu$A. The cusp in Fig. 1(b) corresponds
fairly closely to the CoFeB free-layer's perpendicular saturation
field reflecting the easy-plane anisotropy.

\begin{figure}[h]
\centering
\includegraphics[width=\columnwidth]{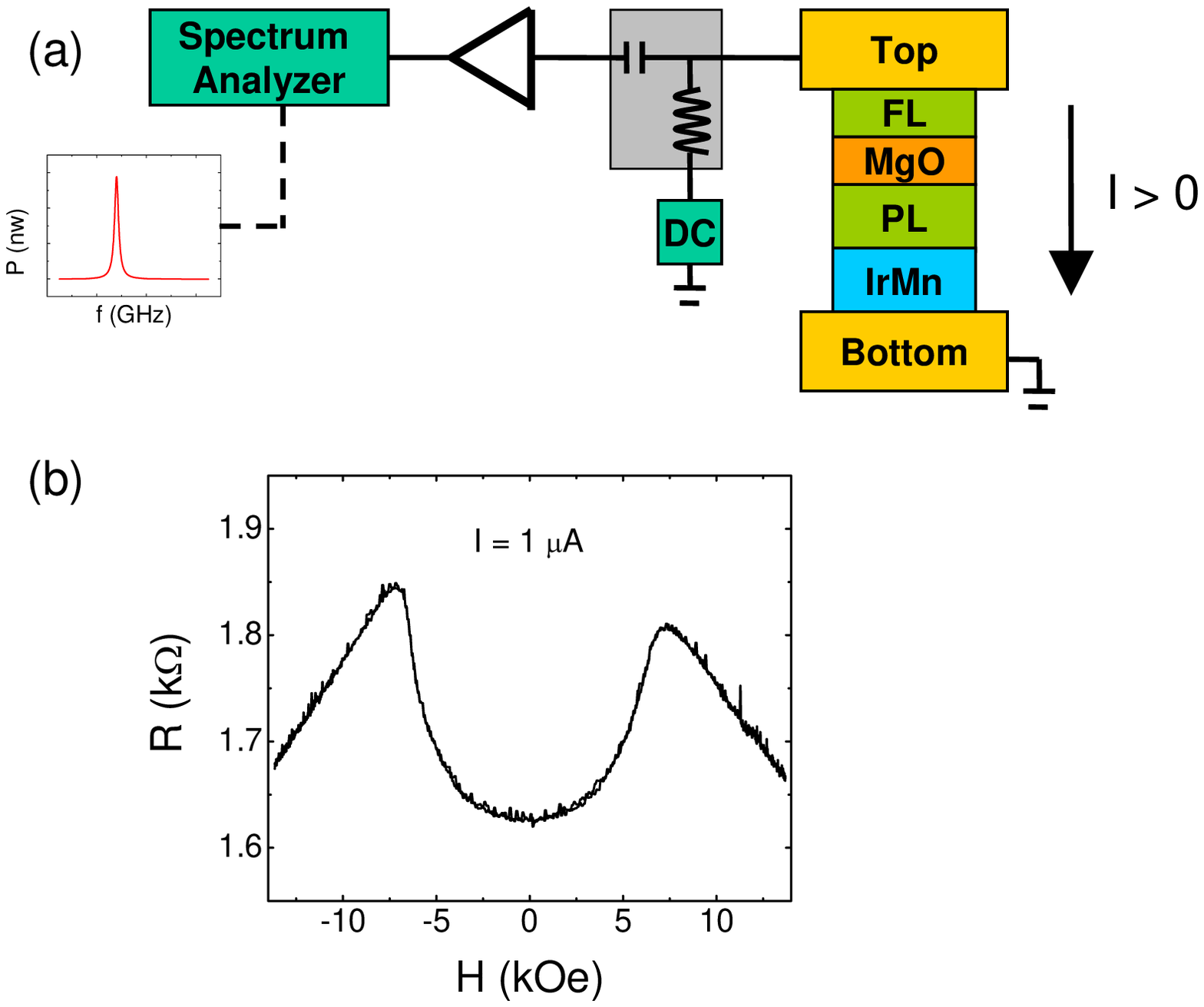}
\caption{(Color online) (a) A simplified diagram of the T-FMR
experimental setup. (b) Perpendicular field dependence of the tunnel
resistance of the MTJ for $I$ = 1 $\mu$A.}
\end{figure}

Figure 2(a) shows typical perpendicular T-FMR spectra of the MTJ
measured at various applied magnetic fields for the same dc bias of
$I$ = 360 $\mu$A. For all data presented, a zero bias spectrum is
subtracted, which removes the background Johnson noise as well as
the amplifier noise. As shown in Fig. 2(a), the T-FMR peaks decrease
in amplitude and shift to higher frequencies with increasing applied
magnetic field.

The perpendicular field dependence of the T-FMR frequency for the
MTJ for $I$ = -360 $\mu$A is presented in Fig. 2(b), where the peak
frequency ($f$) is extracted with a single Lorentzian fit. The data
are well fitted using the perpendicular field Kittel
formula\cite{Kittel}
\begin{equation}
f = {{\gamma}\over{2\pi}}{(H_{res}-4\pi M_{eff})},
\end{equation}
where $\gamma=g\mu_{B}/\hbar$ is the gyromagnetic ratio, $g$ is the
Land$\acute{\textrm{e}}$ factor, and $\mu_{B}$ is the Bohr magneton.
$H_{res}$ denotes the T-FMR resonance field, and $M_{eff}$ denotes
the effective demagnetization. In fitting the data, we treat $g$ and
$M_{eff}$ as free parameters. The best fit gives $4\pi M_{eff}$ =
(6.1 $\pm$ 0.3) kOe and $g$ = 2.0 $\pm$ 0.1 for the 20
$\mathring{\textrm{A}}$ CoFeB free layer of the MTJ. Variations of
$\sim$5$\%$ in $M_{eff}$ and $\sim$10$\%$ in $g$ were observed
between dc bias of -360 and 360 $\mu$A. In addition, $f$ increases
with increasing negative dc bias and exhibits smaller change at
positive dc bias for all the measured values of $H$. This
bias-dependent asymmetry in the T-FMR frequency shift has also been
observed on other MTJs by spin-torque-driven FMR\cite{Kubota-2008,
Sankey-2008}. However, the mechanisms behind it are still unclear so
far.

\begin{figure}[h]
\centering
\includegraphics[width=\columnwidth]{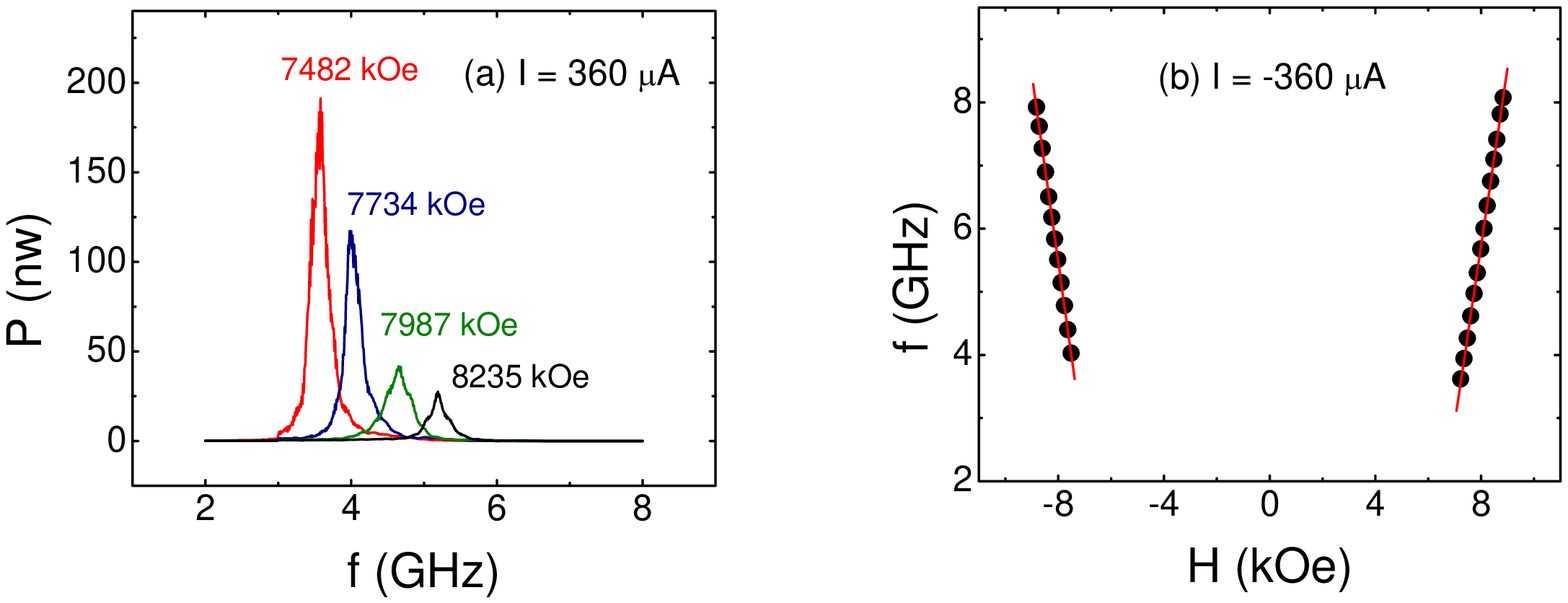}
\caption{(Color online) (a) Typical perpendicular T-FMR spectra of
the MTJ measured at various applied magnetic fields for $I$ = 360
$\mu$A. (b) Perpendicular field dependence of the T-FMR frequency
for the MTJ for $I$ = -360 $\mu$A, including fits to Eq. (1).}
\end{figure}

The $4\pi M_{eff}$ value thus obtained for the 20
$\mathring{\textrm{A}}$ CoFeB free layer of the MTJ is much smaller
than the saturation moment ($4\pi M_{s}$) of bulk CoFeB. This
characteristic is similar to previous reports\cite{Synogatch-2003,
Stutzke-2003, Petit-2007}. For comparison, out-of-plane moments of
extended films with a stack structure of
MgO(30)/CoFeB($t$)/Ta(20)/Ru(20) were measured using a SQUID
magnetometer, where the numbers are layer thickness in angstroms,
CoFeB = Co$_{40}$Fe$_{40}$B$_{20}$, and $t$ = 20 and 100. Before the
SQUID measurements, these extended films were annealed at
300$^{\circ}$C with an in-plane magnetic field of 10 kOe, a
treatment similar to the real stack used for tunnel junction
devices. Figure 3(a) shows the magnetic moment ($4\pi M$) of the
extended CoFeB films as a function of the applied out-of-plane
magnetic field ($H$) for both the thicknesses of 20 and 100
$\mathring{\textrm{A}}$. The saturation moment ($4\pi M_{s}$) of the
CoFeB is determined to be $\sim$15 kOe for $t$ = 100 and $\sim$12
kOe for $t$ = 20. In addition, the saturation field of the CoFeB is
found to be $\sim$6 kOe for $t$ = 20, much smaller than the $4\pi
M_{s}$ value but in reasonable agreement with the $4\pi M_{eff}$
value determined from the perpendicular T-FMR measurements. The
origin of the discrepancy between the measured saturation moment and
the saturation field is not entirely clear. On one hand, the
magnetic granularity of the thin film over a length-scale comparable
to film thickness could result in such reduction of the saturation
field. On the other hand, the possible presence of an
interface-mediated perpendicular magnetic anisotropy could also
reduce the apparent saturation field. Which of these two mechanisms
is responsible for these observations is not yet unambiguously
determined. It should also be noted that a reduced saturation field
is expected in finite size elements due to the exchange and dipolar
contributions to the resonance frequencies\cite{Kalinikos-1986,
Chen-JAP08}. However, since the reference-layer/free-layer (RL/FL)
in the MTJ is nearly compensated and has a thinner total magnetic
thickness, the low-field dipolar correction from the RL/FL in our
case is expected to be much smaller than those discussed for a
spin-valve junction with an uncompensated and much thicker
RL/FL\cite{Chen-JAP08}.

We finally compare the perpendicular T-FMR measurements with the
in-plane T-FMR measurements on the same MTJ. With an in-plane
magnetic field up to 2.5 kOe applied along the hard axis of the
CoFeB free layer, the in-plane T-FMR were measured for various dc
bias up to 270 $\mu$A. Figure 3(b) plots the in-plane hard-axis
field dependence of the T-FMR frequency for $I$ = 270 $\mu$A. The
data are fitted using the Kittel formula for in-plane hard-axis
field\cite{Kittel}
\begin{equation}
f={{\gamma}\over{2\pi}}\sqrt{(H_{eff}-H_{k})(H_{eff}+4\pi M_{eff})},
\end{equation}
where $H_{k}$ is the in-plane uniaxial anisotropy field.
$H_{eff}=H+H_{coupling}+Dk^{2}/{g\mu_{B}}$, where $H_{coupling}$
denotes the effective coupling field between the free layer and the
pinned layer, $D$ is the exchange stiffness, and $k$ is the
spin-wave wave vector. In fitting the data, we take $g$ = 2.0,
neglect the $H_{coupling}$ term, and consider only $k$ = 0 modes
with $H_{k}$ and $M_{eff}$ treated as free parameters. The best fit
gives a $4\pi M_{eff}$ value of (1.1 $\pm$ 0.3) kOe, and a variation
of $\sim$10$\%$ in $M_{eff}$ was observed between dc bias of -270
and 270 $\mu$A. The effective demagnetization field obtained from
the in-plane T-FMR measurements is significantly smaller than those
determined from the perpendicular T-FMR and the out-of-plane SQUID
measurements. This could be due to the possible involvement of
nonmacrospin modes in thermally excited magnetization fluctuations,
specifically the plausible presence of magnetic edge
modes\cite{McMichael-2006, McMichael-2007}. Such modes could make
the quantitative determination of saturation moment and magnetic
anisotropy energy difficult from in-plane T-FMR data.

\begin{figure}[h]
\centering
\includegraphics[width=\columnwidth]{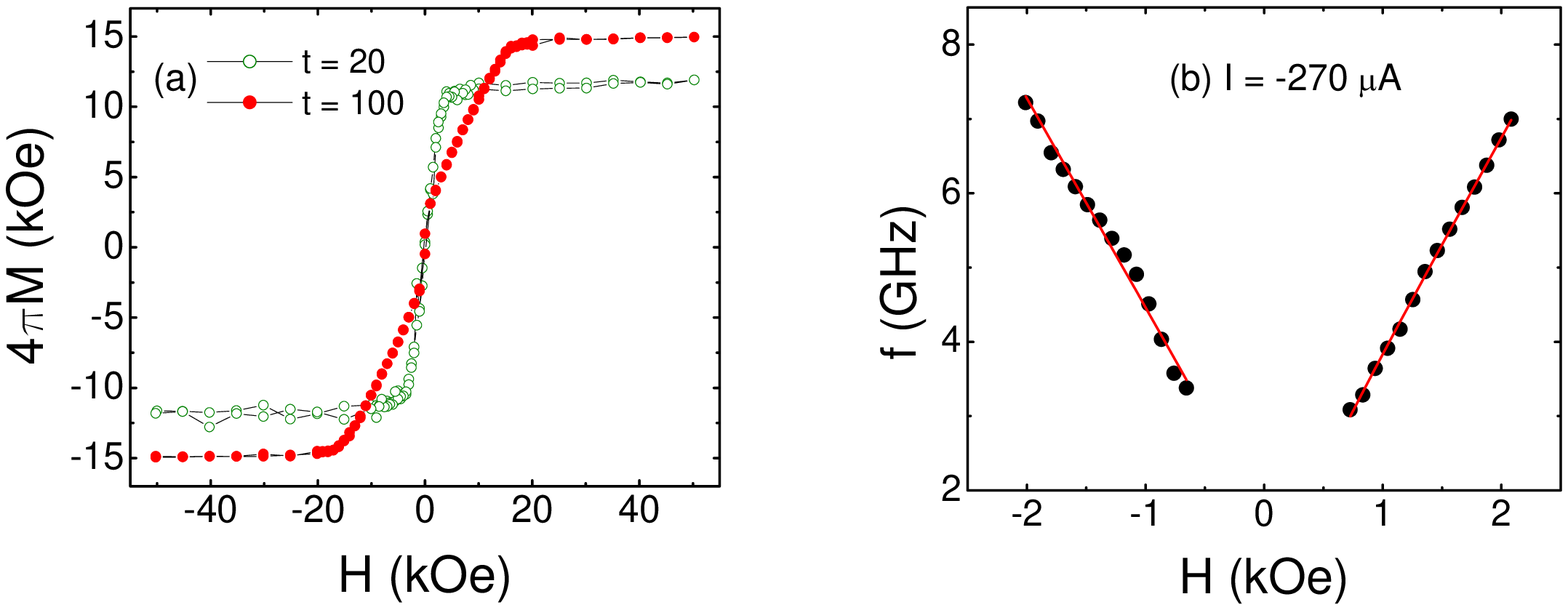}
\caption{(Color online) (a) Out-of-plane SQUID measurements of the
magnetic moment of the extended CoFeB films as a function of the
applied magnetic field for both the thicknesses of 20 and 100
$\mathring{\textrm{A}}$. (b) In-plane hard-axis field dependence of
the T-FMR frequency for the MTJ for $I$ = -270 $\mu$A, including
fits to Eq. (2).}
\end{figure}

In summary, we have found that the perpendicular field dependence of
the T-FMR frequency is in good agreement with Kittel formula. The
extracted effective demagnetization field of the 20
$\mathring{\textrm{A}}$ CoFeB free layer is smaller than the
intrinsic bulk value of CoFeB, but in accord with the saturation
field value determined from the out-of-plane magnetometry
measurements on extended films of the same CoFeB thickness. Our
results suggest that the perpendicular T-FMR is more suitable for
quantitative characterization of the magnetic parameters of
nanoscale MTJs.


We wish to thank the valuable input and support from the MRAM team
at IBM T. J. Watson Research Center. This work also benefited from
an MRAM development alliance program between IBM and MagIC.



\begin{thebibliography}{99}

\bibitem{Sankey-2006} J. C. Sankey, P. M. Braganca, A. G. F. Garcia,
I. N. Krivorotov, B. A. Buhrman, and D. C. Ralph, Phys. Rev. Lett.
\textbf{96}, 227601 (2006).
\bibitem{Fuchs-2007} G. D. Fuchs, J. C. Sankey, V. S. Pribiag, L.
Qian, P. M. Braganca, A. G. F. Garcia, E. M. Ryan, Zhi-Pan Li, O.
Ozatay, D. C. Ralph, and R. A. Buhrman, Appl. Phys. Lett.
\textbf{91}, 062507 (2007).
\bibitem{Chen-2008} W. Chen, J.-M. L. Beaujour, G. de Loubeans, A.
D. Kent, and J. Z. Sun, Appl. Phys. Lett. \textbf{92}, 012507
(2008).
\bibitem{Houssameddine-2008} D. Houssameddine, S. H. Florez, J. A.
Katine, J.-P. Michel, U. Ebels, D. Mauri, O. Ozatay, B. Delaet, B.
Viala, L. Folks, B. D. Terris, and M.-C. Cyrille, Appl. Phys. Lett.
\textbf{93}, 022505 (2008).
\bibitem{Deac-2008} A. M. Deac, A. Fukushima, H. Kubota, H. Machara,
Y. Suzuki, Y. Nagamine, K. Tsunekawa, D. D. Djayaprawira, and N.
Watanabe, Nat. Phys. \textbf{4}, 803 (2008).
\bibitem{Sun-2000} J. Z. Sun, Phys. Rev. B \textbf{62}, 570 (2000).
\bibitem{Nazarov-2002} A. V. Nazarov, H. S. Cho, J. Nowak, S.
Stokes, and N. Tabat, Appl. Phys. Lett. \textbf{81}, 4559 (2002).
\bibitem{Synogatch-2003} V. Synogatch, N. Smith, and J. R. Childress,
J. Appl. Phys. \textbf{93}, 8570 (2003).
\bibitem{Stutzke-2003} N. Stutzke, S. L. Burkett, and S. E. Russek,
Appl. Phys. Lett. \textbf{82}, 91 (2003).
\bibitem{Petit-2007} S. Petit, C. Baraduc, C. Thirion, U. Ebels, Y.
Liu, M. Li, P. Wang, and B. Dieny, Phys. Rev. Lett. \textbf{98},
077203 (2007).
\bibitem{Petit-2008} S. Petit, N. de Mestier, C. Baraduc, C.
Thirion, Y. Liu, M. Li, P. Wang, and B. Dieny, Phys. Rev. B
\textbf{78}, 184420 (2008).
\bibitem{Guan-2009} Y. Guan, D. W. Abraham, M. C. Gaidis, G. Hu, E.
J. O'Sullivan, J. J. Nowak, P. L. Trouilloud, D. C. Worledge, and J.
Z. Sun, J. Appl. Phys. \textbf{105}, 07D127 (2009).
\bibitem{Devolder-2008} T. Devolder, J. Hayakawa, K. Ito, H.
Takahashi, S. Ikeda, P. Crozat, N. Zerounian, J.-V. Kim, C.
Chappert, and H. Ohno, Phys. Rev. Lett. \textbf{100}, 057206 (2008).
\bibitem{Kittel} C. Kittel, {\it Introduction to Solid State
Physics} (Wiley, New York, 2005).
\bibitem{Kubota-2008} H. Kubota, A. Fukushima, K. Yakushiji, T.
Nagahama, S. Yuasa, K. Ando, H. Maehara, Y. Nagamine, K. Tsunekawa,
D. D. Djayaprawira, N. Watanabe, and Y. Suzuki, Nat. Phys.
\textbf{4}, 37 (2008).
\bibitem{Sankey-2008} J. C. Sankey, Y.-T. Cui, J. Z. Sun, J. C.
Slonczewski, R. A. Buhrman, and D. C. Ralph, Nat. Phys. \textbf{4},
67 (2008).
\bibitem{Kalinikos-1986} B. A. Kalinikos and A. N. Slavin, J. Phys.
C \textbf{19}, 7013 (1986).
\bibitem{Chen-JAP08} W. Chen, G. de Loubens, J.-M. L. Beaujour, A.
D. Kent, and J. Z. Sun, J. Appl. Phys. \textbf{103}, 07A502 (2008).
\bibitem{McMichael-2006} R. D. McMichael and B. B. Maranville, Phys.
Rev. B \textbf{74}, 024424 (2006).
\bibitem{McMichael-2007} B. B. Maranville, R. D. McMichael, and D.
W. Abraham, Appl. Phys. Lett. \textbf{90}, 232504 (2007).


\end{thebibliography}
\end{document}